\documentclass{Interspeech}

\interspeechcameraready

\title{DeepFilterGAN: A Full-band Real-time Speech Enhancement System with GAN-based Stochastic Regeneration}

\author[affiliation={1,2}]{Sanberk}{Serbest}
\author[affiliation={2}]{Tijana}{Stojkovic}
\author[affiliation={2}]{Milos}{Cernak}
\author[affiliation={2}]{Andrew}{Harper}

\affiliation{Electrical and Electronics Engineering}{École Polytechnique Fédérale de Lausanne (EPFL)}{Switzerland}
\affiliation{Audio Machine Learning}{Logitech}{Switzerland}
\email{sanberk.serbest@epfl.ch, tstojkovic@logitech.com, mcernak@logitech.com, aharper@logitech.com}

\usepackage{comment}

\begin{document}

\maketitle

\begin{abstract}
    
    In this work, we propose a full-band real-time speech enhancement system with GAN-based stochastic regeneration. Predictive models focus on estimating the mean of the target distribution, whereas generative models aim to learn the full distribution. This behavior of predictive models may lead to over-suppression, i.e. the removal of speech content. In the literature, it was shown that combining a predictive model with a generative one within the stochastic regeneration framework can reduce the distortion in the output. We use this framework to obtain a real-time speech enhancement system. With 3.58M parameters and a low latency, our system is designed for real-time streaming with a lightweight architecture. Experiments show that our system improves over the first stage in terms of NISQA-MOS metric. Finally, through an ablation study, we show the importance of noisy conditioning in our system. We participated in 2025 Urgent Challenge with our model and later made further improvements.
    
\end{abstract}

\section{Introduction}

In the deep learning based speech enhancement (SE) literature, there are many approaches including RNN-based methods \cite{bsrnn}, transformer-based methods \cite{d2former}, Generative Adversarial Network (GAN)-based methods \cite{cmgan} and diffusion-based models \cite{sgmse,storm}. Many works directly focus on denoising or dereverberation tasks \cite{sgmse} while there is also an attempt to obtain a universal SE system that tackles many distortion types together \cite{universe}. While the superiority of such models is an ongoing research question, some advantages and disadvantages of different approaches are reported. The deep learning based SE models can be divided into two main categories, namely predictive and generative models. Predictive models aim to estimate the mean of the clean speech distribution conditioned on the input noisy speech while the generative models try to learn the clean speech distribution instead of its mean. While estimation of the mean in predictive models may cause overdenoising \cite{storm}, the mismatches in the learned clean speech distribution may lead the generative models to hallucinations \cite{schrodinger_nvidia}. On the other hand, thanks to their capability of recovering lost information, the generative structures offer a promising direction for the future of universal speech enhancement. In some works \cite{storm, pfgm, spectraloversubtractionapproachspeech, combinedgenerativepredictivemodeling}, the predictive and generative models are combined in order to utilize the advantages of both approaches. In \cite{storm}, a diffusion based stochastic regeneration framework is proposed. In \cite{pfgm}, the extended Poisson flow generative model is used with a predictive model while \cite{spectraloversubtractionapproachspeech} uses a GAN in a similar setting.

In addition to the output speech quality, another aspect of speech enhancement models is the deployment requirements. For some speech enhancement applications such as mobile telecommunication, the SE system must be able to process streaming data, which requires a real-time and causal system. This requirement may limit the output speech quality because the model can't exploit some temporal relations in the data due to its causal structure. In addition, the real-time requirement imposes a restriction on the network size, which may reduce the representation capacity of the model. The diffusion-based systems \cite{sgmse, storm, universe} usually have a big network size and require several denoising steps, i.e. multiple forward passes, during the inference time. Given that GANs require only a single forward pass during inference and can achieve good performance even with a compact network, they are a good candidate for the real-time generative speech enhancement systems.

In this work, a real-time, low-latency speech enhancement model based on stochastic regeneration with a GAN is proposed. Our system consists of two stages, namely a predictive first stage and a generative second stage. The predictive first stage is adopted from DeepFilterNet \cite{deepfilternet2}. The second stage is a GAN where its generator adopts the network of OnlineSpatialNet \cite{online_spatial_net} and its discriminator adopts the discriminator of MelGAN \cite{melgan}. Our training scheme consists of two steps. The first stage predictive model is trained on the training dataset as a standalone speech enhancement model. In the second step, the intermediate enhanced speech, i.e. the output of the pre-trained first stage, is concatenated with the noisy input speech to provide further noise information and fed into the second stage. During this step, the first stage is kept frozen while the second stage is trained on the same training dataset. Our system processes full-band audio in the time-frequency domain. We show that a GAN based second stage can improve the performance of a predictive first stage without breaking the real time requirements as well as the usefulness of conditioning on the noisy speech.

\section{Background and Related Works}

A conventional single channel speech enhancement system tries to recover the clean speech signal $\mathbf {x(t)}$ by observing the noisy speech signal $\mathbf {y(t)} = \mathbf {h(t)}*\mathbf {x(t)}+\mathbf {n(t)}$ where $\mathbf{h(t)}$ is the impulse response of the channel and $\mathbf {n(t)}$ is an additive noise. This relation can also be represented in short-time Fourier Transform (STFT) domain as $\mathbf {Y(k, f)} = \mathbf {H(k, f)} \cdot \mathbf {X(k, f)} +\mathbf {N(k, f)}$ where $\mathbf {X(k, f)}$ is the STFT of clean speech signal and $\mathbf{k,f}$ being the time and frequency indices. Further distortion types such as clipping, bandwidth limitation, codec distortion, packet loss and wind noise increase the severity of the speech distortion leading to a need for more advanced speech enhancement systems. 

\subsection{DeepFilterNet}

DeepFilterNet \cite{deepfilternet2} is a two-stage speech enhancement model that uses deep filtering. The first stage aims to improve the speech envelope and is performed in equivalent rectangular bandwidth (ERB) domain. Then, the second stage estimates the coefficients of a deep filter to further enhance the periodic structure of the speech in complex domain. Due to its strong performance and real-time suitability, we use it as the predictive model in our system’s first stage.

\subsection{Stochastic Regeneration}

\begin{figure}[ht]
  \centering
  \includegraphics[width=0.8\linewidth]{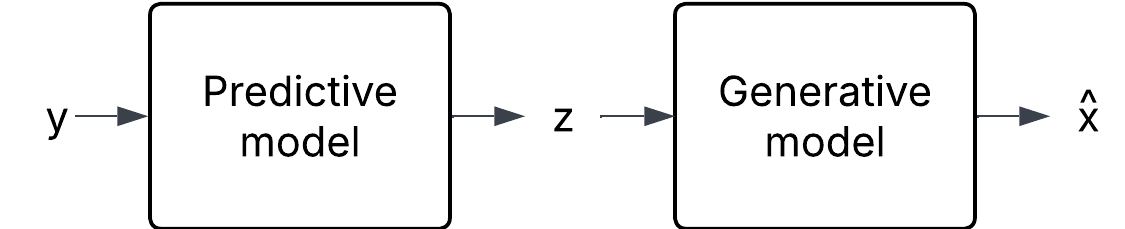}
  \caption{Stochastic regeneration. $y$ is the noisy speech, $z$ is the intermediate enhanced speech and $\hat{x}$ is the final enhanced speech, i.e. the clean speech estimate of the overall system.}
  \label{fig:stochastic_regeneration}
\end{figure}

Due to the nature of their objective functions, the predictive models' output converges to the mean of the clean speech distribution given the noisy input speech, i.e. $\mathbb{E}[\mathbf{x}|\mathbf{y}]$ \cite{storm}. This behavior leads to the removal of fine details in the signal, which corresponds to an overdenoising in speech enhancement. Recovering lost speech content is inherently ill-posed; however, generative approaches can reconstruct finer details in the input signal \cite{storm, pfgm, spectraloversubtractionapproachspeech, combinedgenerativepredictivemodeling}. In the stochastic regeneration framework, the output of a first stage predictive model $\mathbf z$ is fed into the second stage generative model as can be seen in Figure~\ref{fig:stochastic_regeneration}. The output of this second stage is $\mathbf{\hat{x}}$, the estimate of the clean speech.

Different approaches have been investigated in combining the predictive and generative approaches. In \cite{storm}, a stochastic regeneration system is proposed by using a diffusion model for denoising and dereverberation tasks. In \cite{pfgm}, an extended Poisson flow generative model is combined with stochastic regeneration framework built on \cite{storm}. \cite{combinedgenerativepredictivemodeling} introduces a speech super resolution model where they condition a diffusion model on the output of a predictive model. \cite{spectraloversubtractionapproachspeech} uses a similar idea based on GANs to compensate for the spectral oversubtraction in the context of Robot Ego Speech Filtering where they aim to enhance speech detection during human interruptions in the robot's speech.

\subsection{Online SpatialNet}

Online SpatialNet \cite{online_spatial_net} is a multichannel online speech enhancement model for static and moving speakers. It utilizes Mamba \cite{mamba}, an RNN-like structured state space sequence model to store the most relevant historical information. Online SpatialNet uses the spatial information to better enhance the input noisy speech. In our second-stage generator, we modify Online SpatialNet to process both the input noisy speech $\mathbf y$ and the intermediate enhanced speech $\mathbf{z}$ as separate channels. This multichannel-like approach helps retain noise information while also recovering speech components that may have been removed in the first stage.

\subsection{MelGAN}

MelGAN \cite{melgan} is a conditional speech synthesis model based on GANs. It uses a multi-scale architecture comprised of multiple discriminators. Each discriminator focuses on a different frequency range and learns the features in this range. In the second stage discriminator of our proposed model, we adopt the same discriminator structure as in MelGAN. During training, the discriminator learns to distinguish between clean and enhanced speech, while the generator aims to produce enhanced speech that is indistinguishable from clean speech. This adversarial training strategy leads the generator to learn the clean speech distribution.

\section{Proposed Method}

We propose a full-band real-time stochastic regeneration system combining a predictive model with a generative adversarial network (GAN). Our system aims to enhance the performance of the first-stage model, DeepFilterNet2 \cite{deepfilternet2}, by incorporating a GAN-based second stage. The second stage is designed to recover over-suppressed speech segments and further enhance speech quality. The choice of GAN as the generative structure allows for a real-time system.

The first stage enhances the noisy input speech $\mathbf y$ and outputs an intermediate enhanced speech $\mathbf{z}$. This intermediate enhanced speech is already denoised to a good level. However, it may still contain further distortions due to the first stage or some other sources such as clipping, bandwidth limitation etc. The intermediate enhanced speech $\mathbf z$ is then concatenated with the input noisy speech $\mathbf y$ and fed into the second stage, treating them as a two-channel input. The purpose of this operation is two-fold: 

\begin{enumerate}

\item It provides noise information that led to the particular intermediate enhanced speech, which inherently contains a relationship between the noise information and the behavior of the first stage model.

\item It provides the speech content that is possibly removed during the first stage process, which can be useful during the generation.

\end{enumerate}

\begin{figure}[ht]
  \centering
  \includegraphics[width=\linewidth]{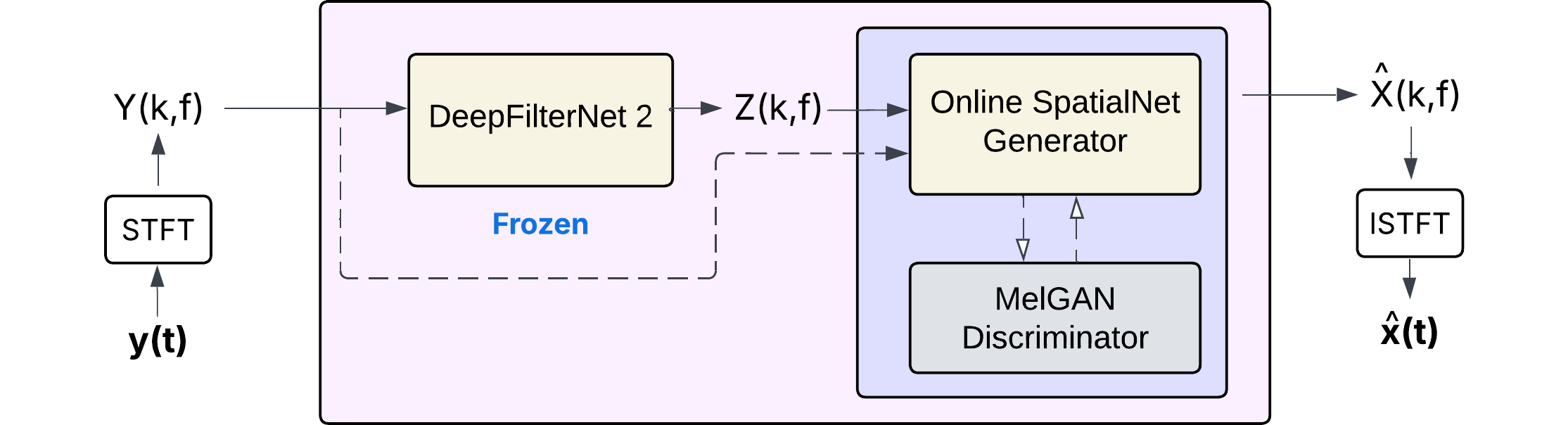}
  \caption{Our proposed model. $Y(k,f)$ is the STFT of noisy speech, $Z(k,f)$ is the STFT of the intermediate enhanced speech and $\hat{X}(k,f)$ is the STFT of the final enhanced speech, i.e. the clean speech estimate of the overall system.}
  \label{fig:deepfiltergan_model}
\end{figure}

As a result, the generator learns the clean speech distribution conditioned on the noisy speech and the intermediate enhanced speech, i.e. $\mathbb{P}(x|y,z)$. Our overall system that processes 48kHz speech samples in STFT domain can be seen in Figure~\ref{fig:deepfiltergan_model}. 

\begin{table*}[t]
  \caption{Results on 2024 URGENT Challenge \cite{urgent} Non-blind Test Set}
  \label{tab:urgent24}
  \centering
  \begin{tabular}{ l c c c c c c c c}

    \toprule
    \bfseries{Model}   &\bfseries{NISQA-MOS $\uparrow$}   &\bfseries{PESQ $\uparrow$} & \bfseries{ESTOI $\uparrow$} &  \bfseries{SDR $\uparrow$}  &\bfseries{LSD $\downarrow$} & 
    \bfseries{Phoneme $\uparrow$} &\bfseries{WAcc \% $\uparrow$} &\bfseries {Overall $\downarrow$} \\ &\bfseries{}&\bfseries{}&\bfseries{}&\bfseries{}&\bfseries{} &\bfseries {Similarity}   

    &\bfseries{}&\bfseries{Ranking} \\
    \midrule
    Noisy &	1.88&	1.41	&0.68&	3.58&	4.89&0.72&	\textbf{79.53}& -\\
    DeepFilterNet2 \cite{deepfilternet2}&	3.28	&2.01	&0.77	&11.01	&4.00&0.78	&74.54 & 	3.06 (4)\\
    UNIVERSE++ \cite{universe} &	\textbf{3.44}&	1.45&	0.60	&5.41&	5.42&0.54	&46.93 &3.63 (5)\\
    First Stage only \\ \textit{(retrained \cite{deepfilternet2})}&	2.66	&\textbf{2.07}&	\textbf{0.80}	&13.00	&4.46&0.79&	75.51 	&2.50 (3)\\
    DeepFilterGAN \\ \textit{without noisy concat} &	2.86&	2.06	&\textbf{0.80}	&\textbf{13.02}	&\textbf{3.67}&0.79&	75.10 &	2.31 (2)\\
    \textbf{DeepFilterGAN with} \\ \textbf{\textit{noisy concat (proposed)}}& 3.12 & 2.03 &	0.79	&\textbf{13.02}	&3.73&\textbf{0.80}&	75.05 & 	\textbf{2.25 (1)}\\
    \bottomrule
  \end{tabular}  
\end{table*}

\subsection{Networks}
The first stage predictive model is DeepFilterNet2 from \cite{deepfilternet2}. It comprises of one encoder and two decoders containing mainly convolutional layers, grouped linear layers and gated recurrent units, adding up to 2.31M parameters. This stage outputs the spectrogram of the intermediate enhanced speech. The second stage is a GAN, i.e. it consists of two networks, namely a generator and a discriminator. The generator network is adopted from \cite{online_spatial_net} by adjusting the number of channels to 2, the number of blocks to 4 and the hidden size to 16, which leads to 1.14M parameters. This network generates a spectrogram conditioned on the intermediate enhanced speech and noisy input speech spectrograms. Finally, the discriminator network in this stage is adopted from \cite{melgan} with the default configuration leading to 0.13M parameters. The discriminator admits the enhanced and clean time signals to differentiate the enhanced samples from the clean ones. The overall system contains 3.58M parameters in total during the training and 3.45M parameters during the inference because the discriminator is only used for training. All elements in our system have low-latency, therefore, it can be used for streaming data applications.

\subsection{Training scheme}
We follow a two-step training scheme. We first train the predictive model independently to stabilize the GAN training. The loss function for this stage combines multiple components:

\begin{enumerate}

\item Spectral loss and multi-resolution spectrogram loss to preserve the frequency structure,

\item Local SNR loss to maintain speech intelligibility,

\item SI-SDR loss to enhance signal quality,

\item Mel-spectrogram $\mathcal{L}_{\text{1}}$ loss to ensure consistency with the clean speech distribution.

\end{enumerate}

We then concatenate the noisy speech spectrogram $\mathbf{Y(k,f)}$ and the intermediate enhanced spectrogram $\mathbf{Z(k,f)}$ obtained from the first stage to feed them into the second stage of our system. During second-stage training, the generator and discriminator compete using a hinge loss objective inspired by \cite{melgan} while first stage model is frozen. The discriminator learns to distinguish real clean speech from enhanced speech, while the generator refines its output to resemble clean speech more closely. The optimization objectives are defined as follows:
\begin{align}
    \min_{D_l}{\mathbb E_x\left [\min(0, 1-D_l(x))\right ] + \mathbb E_{y,z}\left [\min(0, 1+D_l(\hat{x})\right ] }
\end{align}

\noindent $\forall l=1,2,3$ for the discriminator and 
\begin{align}
    \min_{G}{\mathbb E_{y,z}\left [ \sum_{l=1,2,3}{-D_l(\hat{x})}\right ] }
\end{align}

\noindent for the generator where $D_l$ is the $l$th discriminator, $G$ is the generator and $\hat{x}=\textbf{istft}(G(y,z))$, i.e. the inverse STFT of the generator output. In order to push the represented clean speech distribution towards a distribution that contains little distortion, the objective of the generator model is modified to be a weighted sum of the adversarial loss term and a time domain $\mathcal{L}_{\text{1}}$ loss term. The final objective for the generator is as follows:
\begin{align}
    \min_{G}{\mathbb E_{y,z}\left [ \sum_{l=1,2,3}{-D_l(\hat{x})}\right ]  + \beta || x - \hat{x}||_1}
\end{align}
where $\beta$ determines the weighting of the loss terms.
\begin{figure*}[t]
  \centering
  \includegraphics[width=0.8\linewidth]{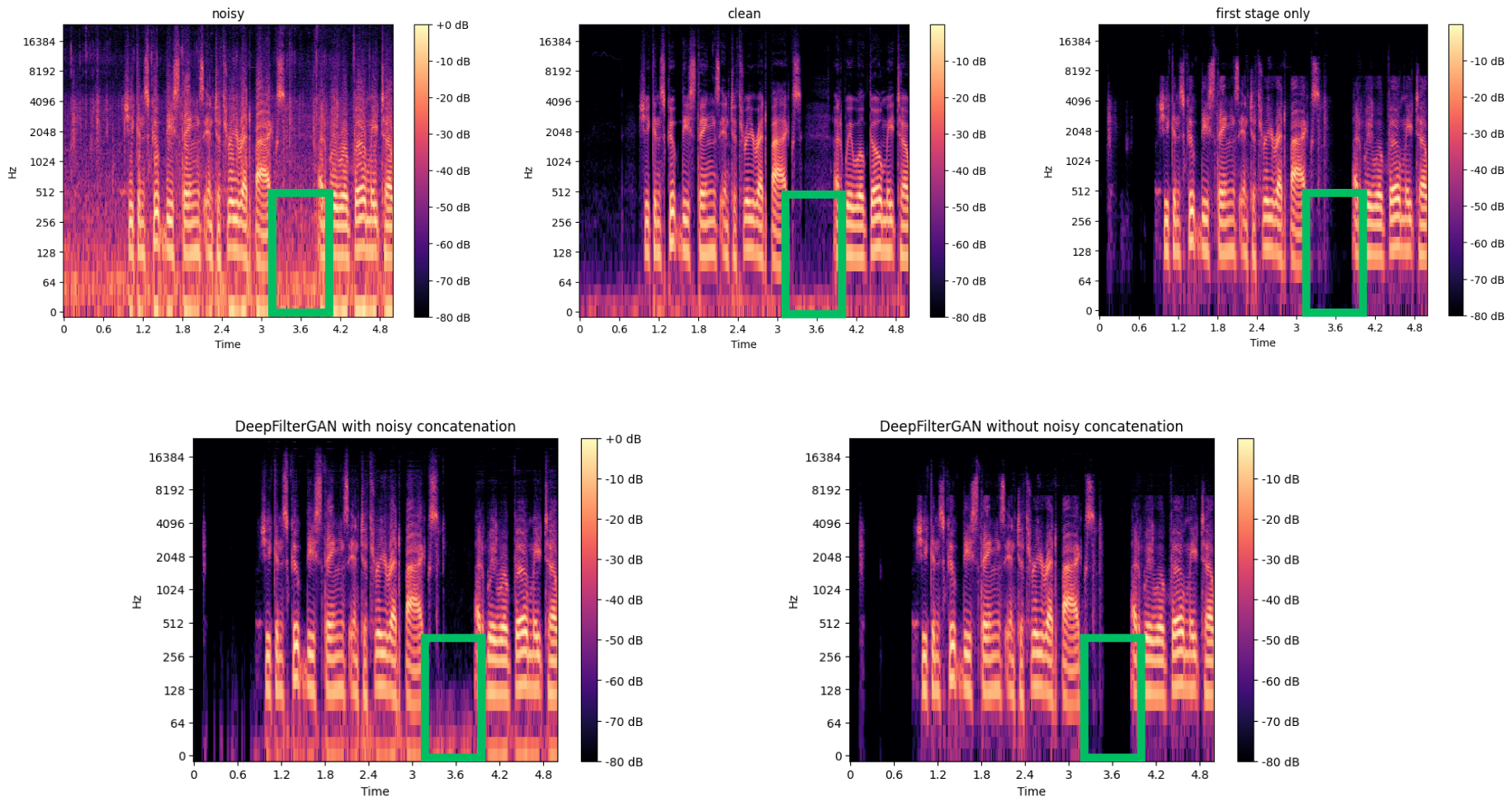}
  \caption{The recovery performance of our proposed system. The area in the green box is removed in the first stage output. Our system with noisy concatenation recovers some portion of this segment while the model without noisy concatenation can't recover it.}
  \label{fig:spec}
\end{figure*}

\section{Experiments and Results}

We trained our proposed system with the training dataset of the 2025 Urgent Challenge \cite{URGENT2025}. This dataset is obtained from a collection of different individual datasets and contains speech samples with various sampling rates, namely 8k, 16k, 22.05k, 24k, 32k, 44.1k, and 48kHz. It covers seven distortion types: additive noise, reverberation, clipping, bandwidth limitation, codec distortion, packet loss and wind noise. 

During training, the samples with different sampling rates are upsampled to 48kHz before the process and downsampled to their original sampling rate after processing. We compute STFT by using a Hanning window with a length of 960 samples (20 ms) and a hop size of 480 samples (10 ms). We use a look-ahead of two frames leading to an overall algorithmic latency of 40 ms. We train the first stage predictive model for 45 epochs. Then, we train the second stage generative part for another 200 epochs while keeping the first stage frozen. We use the same learning rate and weight decay schedule as in \cite{deepfilternet2}. The learning rate schedule starts with a warmup of 3 epochs followed by a cosine decay while the weight decay is controlled by an increasing cosine schedule. We truncate the training samples to 2 s and set the batch size to 64. We also randomly select 180k samples from the whole training set for each epoch which stabilizes the learning procedure.

For a more stable training, we use gradient clipping and weight normalization. The training of the GAN model is a min-max game between the discriminator and the generator. In order to balance the learning progress of these two players, we update the discriminator network every second iteration while we are updating the generator network every iteration. This strategy prevents the discriminator from learning its task, which is easier than that of the generator, too quickly and stabilizes the adversarial training.

Among the related works \cite{storm, universe, pfgm, spectraloversubtractionapproachspeech, combinedgenerativepredictivemodeling}, Storm \cite{storm} and UNIVERSE++ \cite{universe} are the natural choices to compare our model due to the task alignment and code availability. Considering also that UNIVERSE++ \cite{universe} presents superior results over Storm \cite{storm}, we compare our model with UNIVERSE++ \cite{universe} as well as the DeepFilterNet2 \cite{deepfilternet2}, which is the backbone of our first stage. We use the publicly available pre-trained weights for UNIVERSE++ \cite{universe} trained on Voicebank-DEMAND \cite{voicebank_demand} and DeepFilternet2 \cite{deepfilternet2} trained on DNS Challenge \cite{dns4} dataset. We also provide an ablation study to discuss the improvement obtained by the addition of second stage and concatenating the noisy input speech. For the former, we compare the outputs of our model with the outputs of the first stage. For the latter, we trained our system without concatenating the noisy input speech and evaluate the metrics. We use the 2024 Urgent Challenge \cite{urgent} non-blind test data along with the objective evaluation metrics used in the challenge for evaluation of the models. More specifically, we use non-intrusive NISQA-MOS \cite{nisqa} for speech quality estimation, intrusive PESQ \cite{pesq}, SDR \cite{sdr} and LSD \cite{lsd} for noise removal and waveform reconstruction, and intrusive ESTOI \cite{estoi} for speech intelligibility. In addition, we use downstream-task-independent phoneme similarity \cite{phoneme_sim} and downstream-task-dependent word accuracy rate (WAcc) to assess text-to-speech (TTS) and  Automatic Speech Recognition (ASR) performances. We also provide an overall ranking similarly to the Urgent Challenge \cite{URGENT2025}. Firstly, all models are ranked for each individual metric. Then, the metrics within the same category (nonintrusive, intrusive, downstream-task-independent and downstream-task-dependent) are averaged. Finally, the category-based rankings are averaged to obtain overall ranking scores where a lower overall ranking score indicates a superior overall performance. The overall ranking score computation can be summarized as follows:
\begin{align}
    \text{overall score}=\frac{1}{M}\sum_{i=1}^{M}\frac{1}{N_i}\sum_{j=1}^{N_i}\text{ranking}_{ij}
\end{align}
where $M$ is the number of categories, which is 4 in our case, and $N_i$ is the number of metrics within the category $i$. The corresponding results are given in Table~\ref{tab:urgent24}. Our system improves over the first stage in terms of NISQA-MOS without significant degradation in other metrics. One observation is that our system, being a low-latency lightweight model, is left behind the UNIVERSE++ \cite{universe} model in terms of NISQA-MOS. However, the phoneme similarity and word accuracy rate (WAcc) performance of our system is better than that of UNIVERSE++ \cite{universe} indicating better intelligibility and a better ASR performance. Besides, our model achieves the best overall ranking indicating that our model demonstrates a more uniform performance across different metrics compared to other models. Considering that UNIVERSE++ \cite{universe} has 107.5M parameters and includes a diffusion process requiring multiple inference steps and therefore a long inference time, this performance of our 3.58M parameter low-latency system is significant.

Next, we examine the recovery capability of our model for the segments that are removed in the first stage predictive model due to over-suppression. Figure~\ref{fig:spec} shows an example of the outputs for two stages of our model with noisy concatenation. A speech content contained in the ground truth is removed in the first stage indicating an over-suppression. It is observed that our model with noisy concatenation can recover some portion of this speech content. However, if we remove the noisy concatenation from our system, this recovery performance is severely degraded, which shows the importance of this operation.

\section{Conclusion}

In this work, we propose a full-band real-time stochastic regeneration system combining a predictive model with a GAN called DeepFilterGAN. The generative capabilities of the second stage and conditioning on the noisy speech along with the output of the first stage recover the over-suppressed speech content. The choice of a GAN-based generative structure allows for a compact model with 3.58M parameters and a fast processing. The experiments show that our system improves over the first stage predictive model in terms of speech quality. Although it doesn't reach the NISQA-MOS performance of the compared model with high number of parameters, its overall ranking score indicates that the proposed system is promising for streaming data applications. The presented ablation study reflects the importance of conditioning on the noisy speech, which is introduced as a mechanism to provide further noise information to the second stage. Investigation of training both stages together and the performance of the \textit{Mamba block} in the second stage are possible future directions for our proposed system.

\ifinterspeechfinal
\else
\fi

\bibliographystyle{IEEEtran}
\bibliography{mybib}

\end{document}